\documentclass[11pt]{amsart}
\usepackage{geometry}                
\usepackage{graphicx}
\usepackage{amssymb}
\usepackage{amsmath}
\usepackage{epstopdf}
\usepackage{float}
\title{Application of Lossless Data Compression techniques to Radio Astronomy Data Flows}
\author{Tim Natusch}
\date{\today}                                           

\begin{document}

\maketitle
\tableofcontents{}
\section{Introduction}
The modern practice of Radio Astronomy is characterized by extremes of data volume and rates, principally because of the direct relationship between the signal to noise ratio that can be achieved and the need to Nyquist sample the RF bandwidth necessary by way of support. The transport of these data flows is costly. By examining the statistical nature of typical data flows and applying well known techniques from the field of Information Theory the following work shows that lossless compression of typical radio astronomy data flows is in theory possible. The key parameter in determining the degree of compression possible is the standard deviation of the data. The practical application of compression could prove beneficial in reducing the costs of data transport and (arguably) storage for new generation instruments such as the Square Kilometer Array.


\section{Information theoretic basis of data compression techniques}
The fundamental theory of data compression has been well understood (at least) since the publication in 1948 of Claude Shannon's seminal paper "A Mathematical Theory of Communication" \cite{Shannon}. Following the broad intent of his line of reasoning we start with a simple model of a system consisting of a Data Source, a communications Channel and a Data Sink. The Source outputs a discrete stream of data symbols that are conveyed in the form of signals over the Channel to the Sink.

\begin{figure}[H]
\centering
\includegraphics[width=0.8\textwidth]{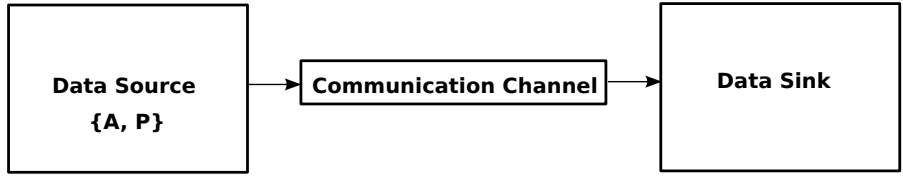}
\caption{Simple communications system model}
\end{figure}

We assume that the data source has a finite set of $N$ symbols that it can generate and send to the communication channel, we refer to this as the Source Alphabet $A=\{a_1,a_2, \dotsc, a_N \}, a_i=i$'th symbol. We further assume that the source randomly generates symbols from the set $A$ and that the generation of any given symbol is independent of any previous activity of the source (we restrict ourselves to consideration of ``memoryless'' systems only to begin with; the conditional probability that symbol $a_j$ is emitted following symbol $a_i$ is $P(a_j | a_i))=P(a_j)$ i.e. the symbols are statistically independent). The combination of Alphabet $A$, the set of probabilities for symbol generation by the source $P=\{P_{a_1},P_{a_2},\dotsc,P_{a_N}\}\:\:(P_{a_i}=$ probability that the source emits the i'th symbol) and knowledge of the transmission bandwidth $W$ of the channel is sufficient to completely characterise the behaviour of this system for our current purpose. A more complete schematic of a system would include Source Coder and Source Decoder blocks between the Source and Channel, and the Channel and Sink respectively. These blocks convert symbols to and from physical signals of type appropriate for transmission via the particular channel media (e.g. twisted pair, fibre optic cable, \dots) used. These additional components have no bearing on the discussion here, further mention is accordingly omitted. \\

There are a number of fundamental assumptions that underpin Information Theory \cite{Shannon}\cite{url1}; \\

\begin{itemize}
  \item Information is gained by the Sink whenever the Source emits a symbol.
  \item The amount of information gained is a function of the probability associated with emission of the symbol ($P_{a_i}$).
  \item The less probable the emission of a symbol the greater the amount of information gained when it is emitted. 
\end{itemize}

\[I(a_i)=\log\left(\frac{1}{P(a_i)}\right)=-\log(P(a_i))\] is the normal working definition adopted to quantify the amount of information gained when the source emits symbol $a_i$. Where the base of the logarithm is $2$ information is quantified in units of ``bits'';

\[I(a_i)=-\log_2(P(a_i))\:\text{bits}\] 

\section{Data Compression}
If a source emit symbols at a fixed rate $R_{source}$ [symbols/unit time] then in a time $T$ the source will emit a total of $N=R_{source}\times T$ symbols. At its most basic data compression is the act of removing some subset of the emitted symbols. If $n$ symbols are removed from the source output data flow during interval $T$ then the remaining set of $N-n=L$ symbols represent a ``compressed'' version of the original data flow. It is useful at this juncture to define a compression ratio \[C_R=\frac{N-n}{N}=\frac{L}{N}\] (which is clearly $<1$ for $n > 0$). If the channel continues to transmit symbols at the rate $R_{source}$ then the compressed data can be conveyed to the Sink in a shorter interval of time $T'=C_R\times T$. \\\\Alternatively the compressed data can be transferred across the channel at a lower (compressed) rate in the original transfer interval $T$; \[R_{compressed}=\frac{N-n}{T}=\frac{L}{T}=\frac{N\times C_R}{T}=\frac{R_{source}\times T\times C_R}{T}=R_{source}\times C_R\] \\Compression has the advantages of reducing the time required to transmit a fixed block of data or to allow for lower speed transfer over a bandwidth limited channel in a given quantum of time. \\\\The ``universal law'' of nature stating that ``you don't get anything for nothing'' would suggest that with the removal of symbols from the data flow the Sink inevitably losses information. This need not necessarily be so as (dependent upon the means of compression ) it may be possible to completely recover the original data from its compressed version. Accordingly we may define two classes of compression algorithm; ``lossy'' and ``lossless''. In the former case the information loss is irretrievable and of use only if we can accept this (as for instance is the case with JPEG compression of images, high spatial frequency content, the fine detail, being discarded). In the latter case of lossless compression some algorithm for ``decompression'' that recovers the original data in entirety exists. In what follows the discussion is restricted solely to lossless compression techniques.

\begin{figure}[H]
\centering
\includegraphics[width=0.8\textwidth]{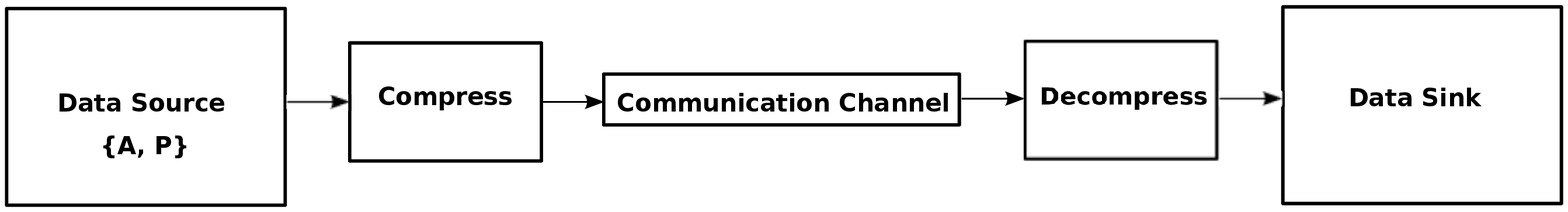}
\caption{Lossless Compression model}
\end{figure}

\section{Source Entropy}
A readily computed property of the source is its ``First Order Entropy'';
\[H(A,P)=\sum_{i=1}^{N}P(a_i)I(a_i)=-\sum_{i=1}^{N}P(a_i)\log_2(P(a_i))\]
which examination reveals to be nothing more than the average number of bits of information obtained by observing the source emit some arbitrary symbol. The relevance of this parameter is that it represents a lower bound for lossless compression in accordance with the Lossless Source Coding Theorem of Shannon \cite{Shannon}\cite{url1};\\
\[H(A,P) \leq L \leq H(A,P)+\frac{1}{N}\] \\ where $N=$ length of the uncompressed data string and $L=$ the length of the compressed string. For very long sequences of symbols $N\rightarrow \infty$ and $H(A,P)+\frac{1}{N} \rightarrow H(A,P)$ squeezing $L$ to be\\ \[L=H(A,P)\]\\

Let some arbitrary source have an entropy of $H\:bits$ per symbol then Information Theory tells us that if we are transmitting data over the channel using (say) $X\:\text{bits}$ per symbol (where $X > H$) that there is ``redundancy'' of $R_D=X-H\:\:\text{bits}$ in the data. This redundancy may be exploited to achieve lossless compression as long as the compression ratio used $R_C$ remains such that\\ \[R_C \geq \frac{H}{X}=\frac{X-R_D}{X}=1-\frac{R_D}{X}\] \\

Two other useful metrics are the fractional compressibility \[C_{frac}=1-R_C=1-(1-\frac{R_D}{X})=\frac{R_D}{X}\] 
and the percentage compressibility\[C\:[\%]=100 \times C_{frac}\]\\

Note that whilst Information Theory tells us this that if the conditions above are adhered to then lossless compression is possible it sadly does not of itself inform us of how this is to be accomplished, other ingenuity must be exercised to arrive at suitable algorithms. Fortunately considerable effort has been expended on this and there are a variety of suitable algorithms for us to choose from, see \cite{alg1} for example.\\

Source Entropy is maximized when the source symbols are emitted with equal probability. This is most easily demonstrated in the case of a source with just two symbols in its alphabet, say $A=\{0, 1\}$. Let $P(0)$ be the probability that the source emits a $0$ and $P(1)$ the probability $1$ is emitted. Adding the constraints that the sum of the probabilities must be unity (the source is certain to emit some symbol) and $0\leq P(i) \leq 1$  we then have\\
\\
\[P(0)+P(1)=1 \implies P(1)=1-P(0)\]\\
\\
and the entropy can then be written as;\\
\\
\[H=-P(0)\log_2(P(0))-(1-P(0))\log_2(1-P(0))\]\\
\\
Plotting this for all $P(0)\in [0,1]$ (the allowable values) we readily see that entropy is at a maximum when the symbols are equiprobable i.e $P(0)=P(1)=0.5$.
\begin{figure}[H]
\centering
\includegraphics[width=0.8\textwidth]{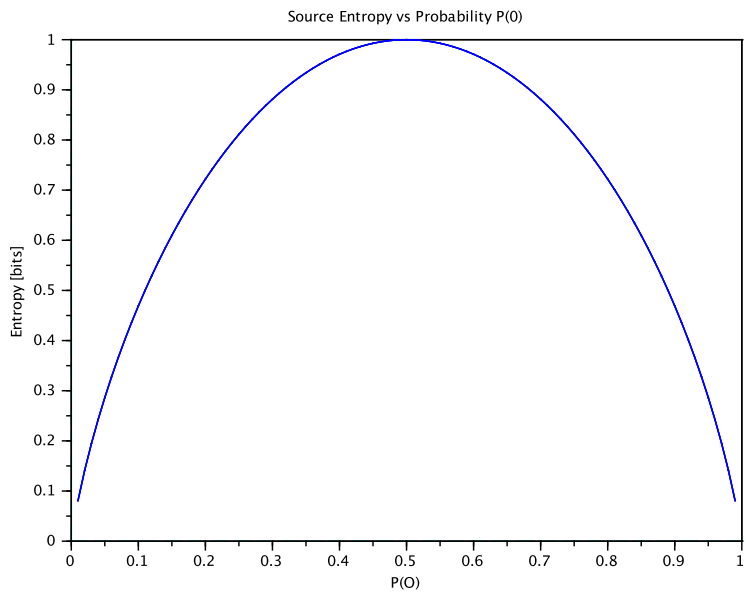}
\caption{1 bit entropy curve}
\end{figure}

A corollary of this is of course that entropy is reduced, the redundancy increases and the amount of compression that can be achieved increased whenever the symbol distribution deviates from the equiprobable state.

\section{The nature of radio sources and the output response of a receiving system}

The signals received from a typical radio astronomical source are assumed to be the result of superposition of emissions from an extremely large number of independent oscillators. As a consequence the distribution of signal voltages intercepted by the antenna of a radio telescope is random and typically well modeled by either a continuous Gaussian or Raleigh probability distribution \cite{Burke} \cite{Rolfs} . Let us settle on a continuous Gaussian probability distribution\\ \\ \[P(x,\mu,\sigma)=\frac{1}{\sigma \sqrt{2\pi}}e^{\left(\frac{-(x-\mu)^2}{2\sigma^2}\right)}\]\\ \\ $\mu=$ average value \\ $\sigma=$ standard deviation \\ \\ with mean value $\mu=0$ to describe the statistical nature of the observed signal. \\
\\
This signal is then amplified, filtered and typically frequency shifted (heterodyned) before finally being digitized by the components of a receiving system on the downstream side of the antenna. Amplification, filtering and mixing have minimal effect on the statistical properties of the signal.\\
\begin{figure}[H]
\centering
\includegraphics[width=0.8\textwidth]{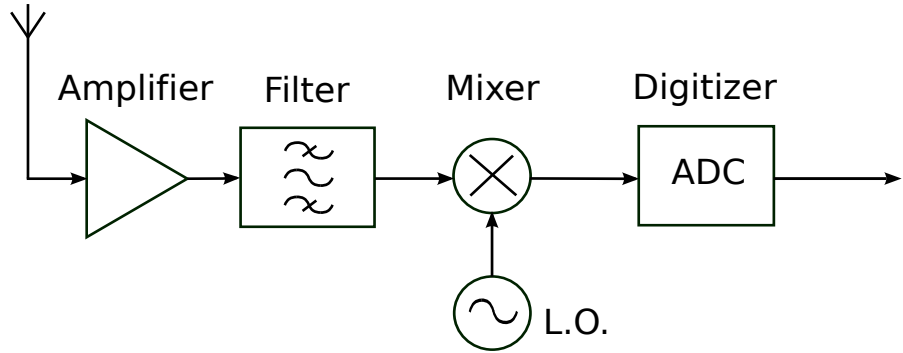}
\caption{Representative signal processing and flow in a typical receiving system}
\end{figure}

The last processing function (Analogue to Digital Conversion) discretizes / quantizes the analogue signal (introducing quantization noise in the process) and (typically) shifts the mean signal level to some non zero value. Usually the mean value will be set close to $\mu=(2^N-1)/2$ (the mid point of the ADC quantization range) where $N=$ no. of bits in the output word of the ADC.\\
\\
\begin{figure}[H]
\centering
\includegraphics[width=0.8\textwidth]{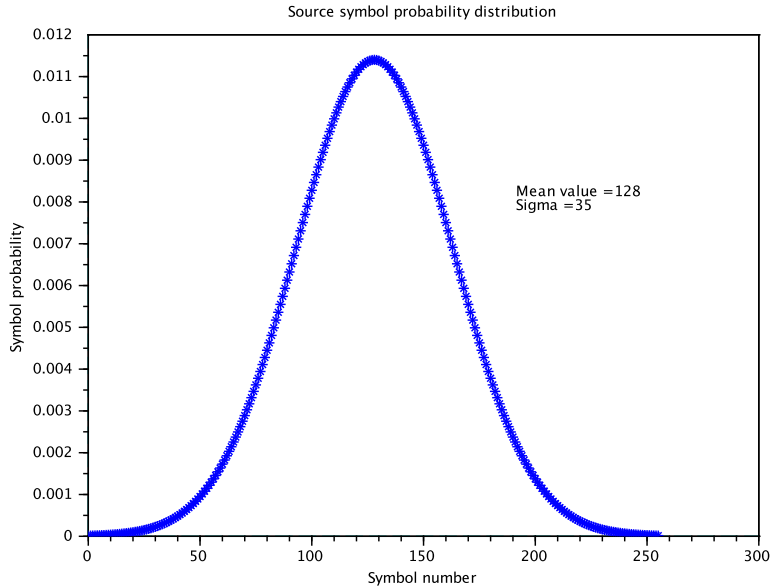}
\caption{Example normalised Gaussian distribution; mean value 128, sigma 35}
\end{figure}

The output of the ADC, the ``source'' that any downstream compressor sees is then a stream of symbols chosen from the finite set $A=\{0,1,...,2^N-1\}$ with discrete Gaussian probability distribution of mean value $\mu \approx (2^N-1)/2$. Crucially, note that the symbols emitted by the source (ADC output) are far from equiprobable.

\section{The case of 2 bit VLBI data streams} 

It is common practice in VLBI to use 2 bit sampling, the source then has an alphabet consisting of $2^2=4$ symbols; $A=\{00,01,10,11\}$. In order to minimise the effects on cross correlation amplitude some effort is expended through use of AGC (Automatic Gain Control) and other means to maintain an optimum distribution of bit statistics; typically $P=\{0.33,0.17,0.17,0.33\}$ \cite{TMS-1}\cite{mk5memo-1}.
The first order source entropy is then;\\
\\
\[H(A,P)=-0.33\log_2(0.33)-0.17\log_2(0.17)-0.17\log_2(0.17)-0.33\log_2(0.33)\approx 1.925\:[bits]\]
\\
Such a 2 bit stream is then ``losslessly' compressible by the factor $R_C\approx (2-1.925)/2\approx 0.0375$.\\

Whilst compression by some $3.75\:\%$ is in theory achievable for this common case, the effort / cost of compressing for such a low return is unlikely to be regarded as worthwhile.
 
 \section{8 bit sampled data streams}
In the case of the use of 8 bit Analogue to Digital Converters (ADC) to digitize the analogue signal from a radio astronomy receiver the data source will have an alphabet with a total of $2^8=256$ symbols; \\
\\
$A=\{00000000, 00000001, 0000010, ... , 11111110, 11111111\}_b=\{0,1,2,...,255\}_{10}$\\
\\
To compute source entropy we require knowledge of the symbol probability distribution $P$. As justified above let us assume a Gaussian distribution and additionally a mean value of $128$ ($=2^7$), the mid-point of the 8 bit ADC range (rounded to the nearest whole number). We may then readily determine by direct computation source entropy for a range of different values of standard deviation. This has been done using Scilab code, results are plotted in the figures that follow. By inspection we see that source entropy falls (figure 6), the redundancy (figure 7) and the scope for compression (figure 8) increase as the standard deviation reduces.\\
\\

\begin{figure}[H]
\centering
\includegraphics[width=1.0\textwidth]{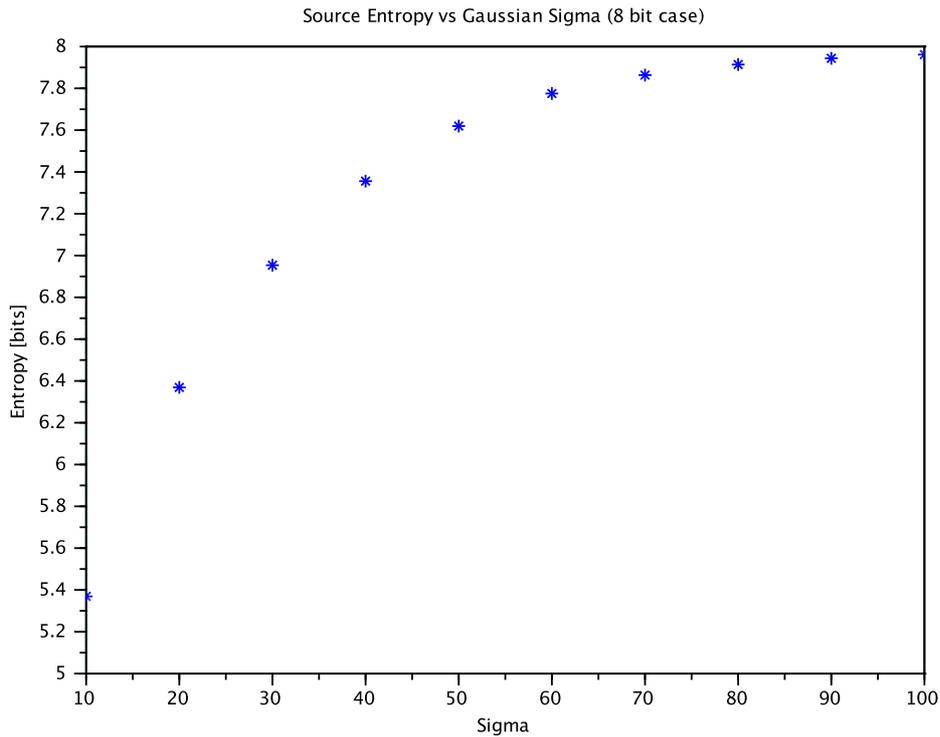}
\caption{Variation of source entropy with Gaussian Sigma}
\end{figure}
\begin{figure}[H]
\centering
\includegraphics[width=1.0\textwidth]{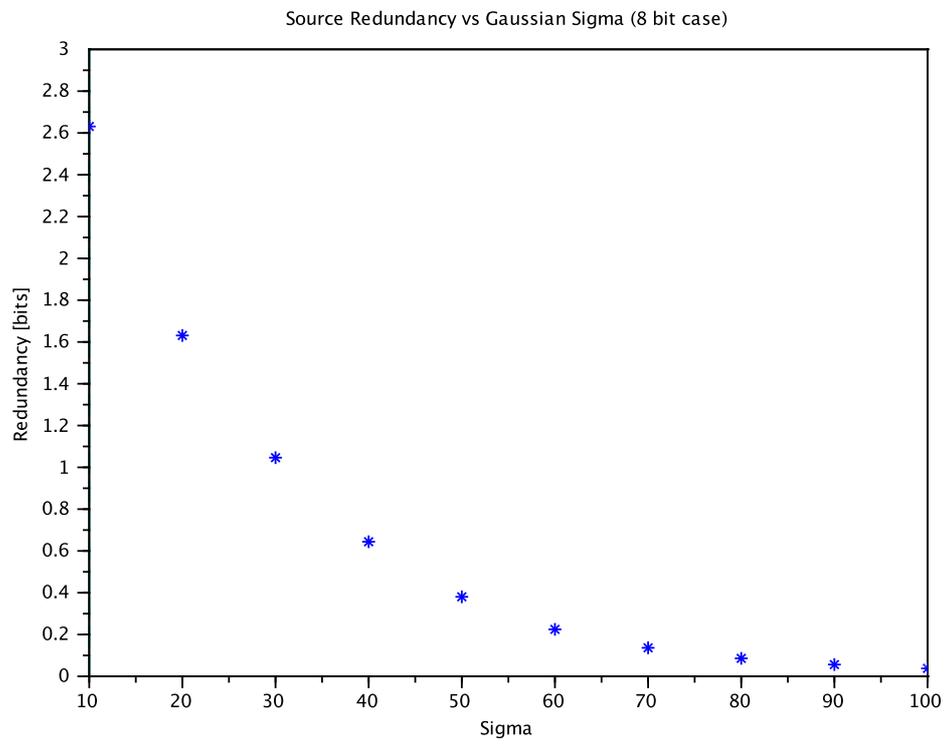}
\caption{Redundancy vs Sigma [bits]}
\end{figure}
\begin{figure}[H]
\centering
\includegraphics[width=1.0\textwidth]{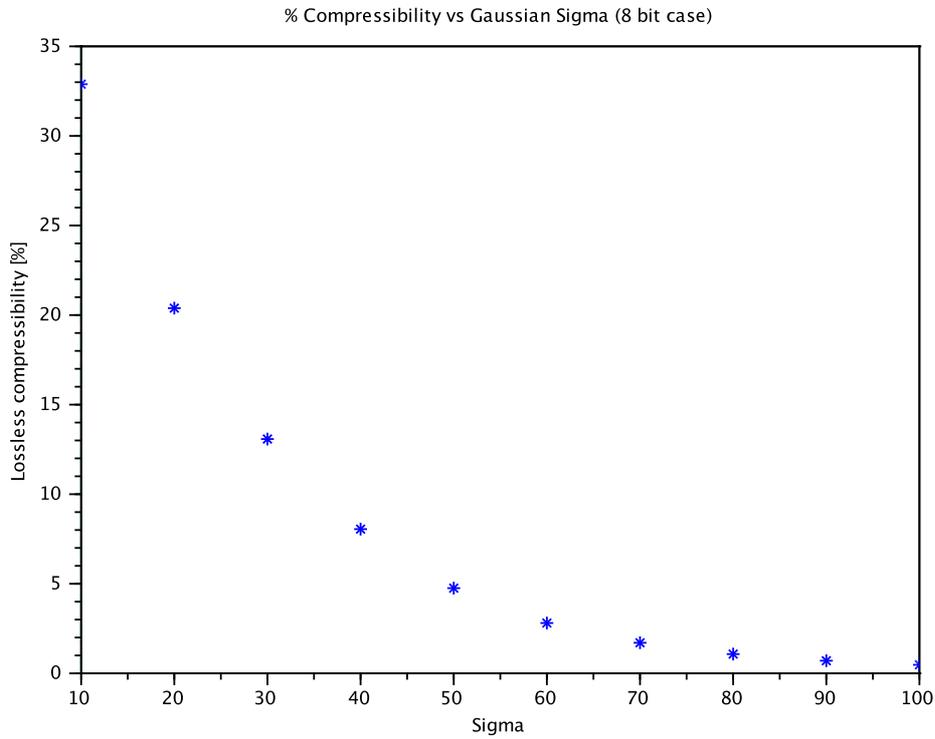}
\caption{Compressibility vs Sigma [\%]}
\end{figure}
As a matter of practice it is the case that the fractional and \% Compressibility curves for the 8 bit case (within the range of sigma examined) are found to well modeled by a negative exponential function as a result of performing a Least Squares fit. In particular;
\[C[\%]=53.081e^{-0.048\sigma}\] with a Coefficient of Determination $R^2=0.99878$. This does not of course constitute a proof that this is the correct relationship but is readily calculated and arguably is sufficiently accurate for practical work with the 8 bit case.
\begin{figure}[H]
\centering
\includegraphics[width=1.0\textwidth]{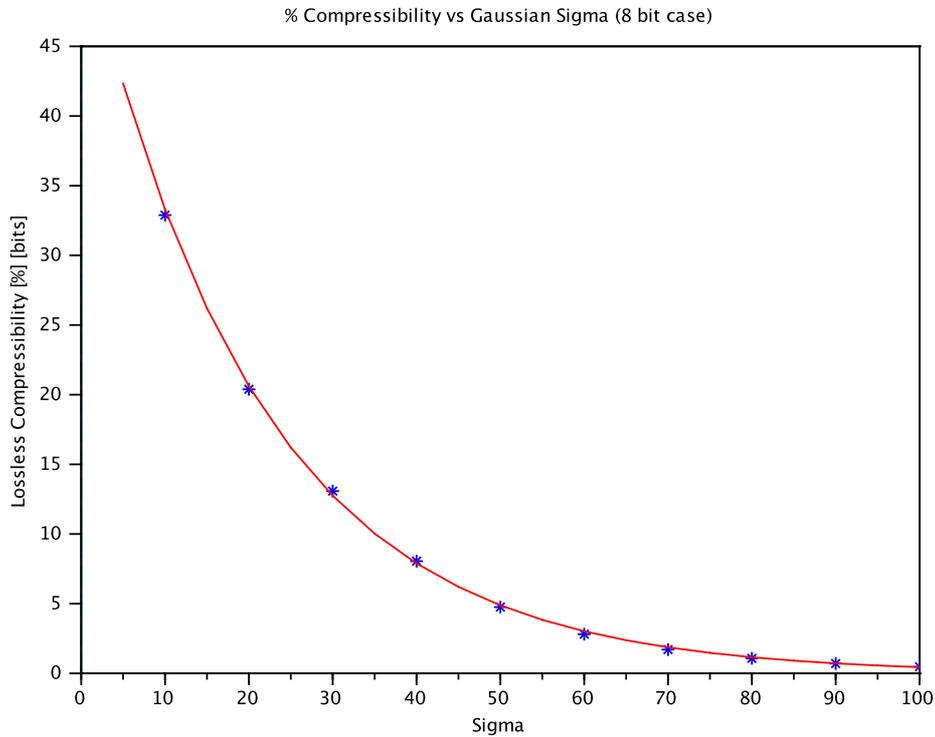}
\caption{Exponential fit to C[\%] : $R^2=0.99878$}
\end{figure}

\section{8 bit test cases}
%

Several data files recorded with 8 bit quantization with equipment set up for the then experimental New Zealand station BART6 \cite{Bart6} have been subject to analysis. The BART 6 station used a MAX1448 10 bit ADC with just the 8 most significant bits fed onto the input of a PCEVN recorder system \cite{url2}\cite{url3} and used recording software fauto \cite{Philips} to control the process and deliver files formatted for use with the correlator of the Australian Long Baseline Array (LBA) \cite{lba-1}. The statistical distribution of the first $5$ million data points from these files was determined using the Scilab code of appendix B. Note that limitations of stacksize prevented the Scilab code from opening the entire file on the MAC OS X platform used for this analysis. 
\begin{table}[htdp]
\caption{Data file statistics and entropy}
\begin{center}
\begin{tabular}{|c|c|c|c|c|}\hline 
File name &  File size, Bytes & Mean value & $\sigma$ & Entropy H, bits\\ \hline \hline
 NZ\_vt03e\_069\_225750.lba & 320,000,016 &$\approx 127.72$ & $\approx 9.70$ & $\approx 5.33$ \\ \hline
 NZ\_ikaros\_197\_123005.lba  & 640,000,016 & $\approx 127.69$ & $\approx 16.73$ & $\approx 6.11$ \\ \hline
\end{tabular}
\end{center}
\label{Compression results}
\end{table}

\begin{figure}[H]
\centering
\includegraphics[width=1.0\textwidth]{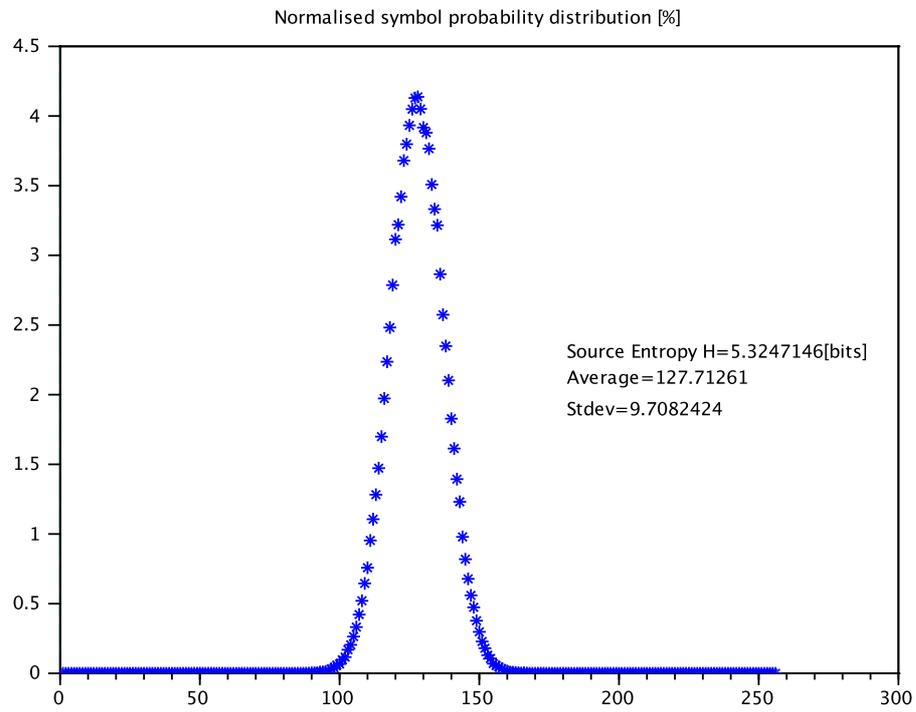}

\caption{Symbol probability distribution of file NZ\_vt03e\_069\_225750.lba}
\end{figure}
\begin{figure}[H]
\centering
\includegraphics[width=1.0\textwidth]{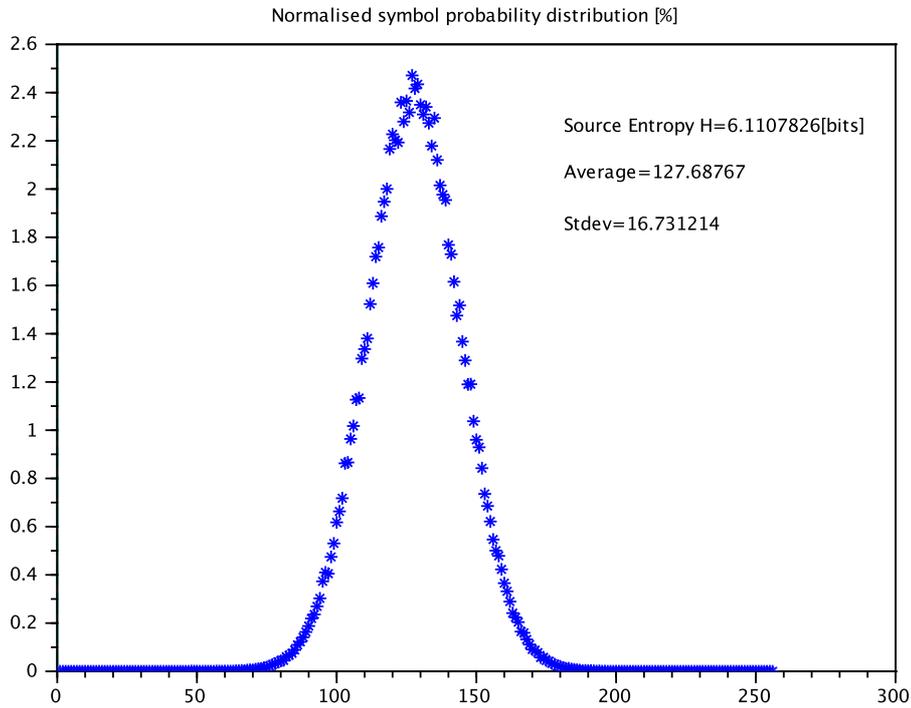}

\caption{Symbol probability distribution of file NZ\_ikaros\_197\_123005.lba}
\end{figure}

The theoretical redundancy and compressibility of these two files are;
\begin{table}[htdp]
\caption{Entropy and theoretical \%compressibility}
\begin{center}
\begin{tabular}{|c|c|c|}\hline 
File name  & Entropy H, bits & C \%\\ \hline \hline
 NZ\_vt03e\_069\_225750.lba & $\approx 5.325$ & $\approx 33 $ \\ \hline
 NZ\_ikaros\_197\_123005.lba  & $\approx 6.11$ & $\approx 23.6$ \\ \hline
\end{tabular}
\end{center}
\label{Theoretical compressibility}
\end{table}
\\
The results of compressing these files with a variety of different algorithms available from the command line of an OS X based computing platform are displayed in Tables 3 and 4. All compressed files were decompressed and checked for errors against a copy of the original file with the OS X "diff" command to verify the decompression.\\
\\

\begin{table}[htdp]
\caption{Compression results for NZ\_vt03e\_069\_225750.lba}
\begin{center}
\begin{tabular}{|c|c|c|c|c|}\hline 
Algorithm & Compressed file size, Bytes & C \% & Decompression verified\\ \hline \hline
gzip & $223,006,066$ & $\approx 30.3$ & $\checkmark$ \\ \hline
bzip2 & $222,821,526$ & $\approx 30.3$ & $\checkmark$ \\ \hline
lzma & $218,170,285$ & $\approx 31.8$ & $\checkmark$ \\ \hline
\end{tabular}
\end{center}
\label{Compression results}
\end{table}
\begin{table}[htdp]
\caption{Compression results for NZ\_ikaros\_197\_123005.lba}
\begin{center}
\begin{tabular}{|c|c|c|c|c|}\hline 
Algorithm  & Compressed file size, Bytes & C \% & Decompression verified\\ \hline \hline
gzip & $496,895,469$ & $\approx 22.3 $ & $\checkmark$ \\ \hline
bzip2 & $512,216,439$ & $\approx 19.9 $ & $\checkmark$ \\ \hline
lzma & $498,010,177$ & $\approx 22.2 $ & $\checkmark$ \\ \hline
\end{tabular}
\end{center}
\label{Compression results}
\end{table}


These results are plotted against the maximum compressibility possible in Figure 12 and lend support to an assertion of this work, that the compressibility scales with the standard deviation of the file. More robust support would of course be possible if data with a wider range of standard deviation ($\sigma$) were available but at the time of writing the equipment used to generate 8 bit files has been decommissioned. 
\\
\begin{figure}[H]
\centering
\includegraphics[width=1.0\textwidth]{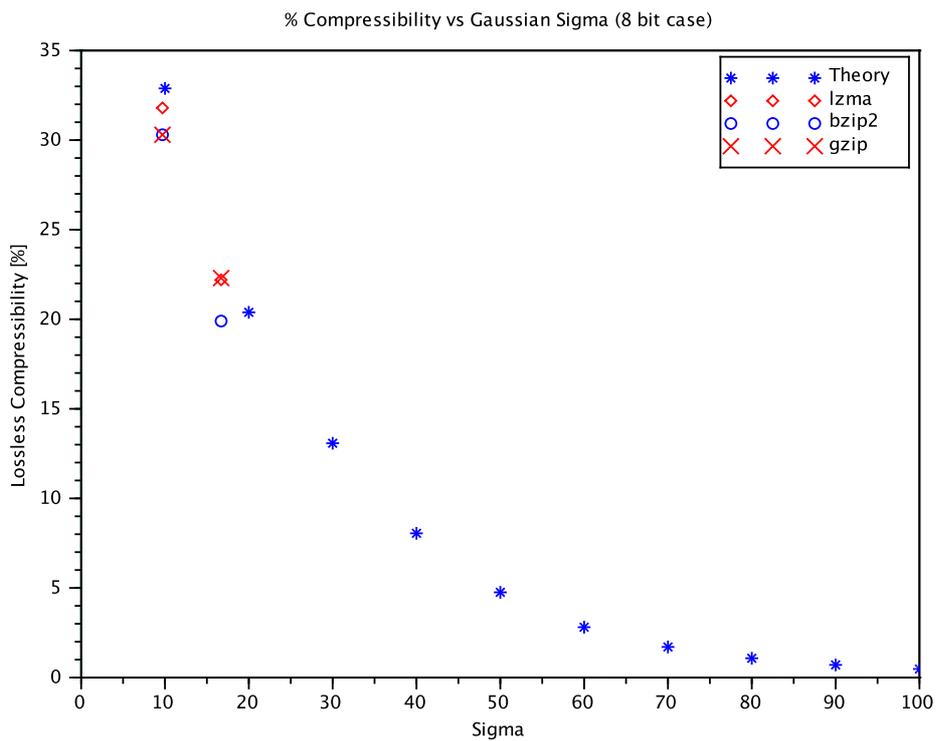}

\caption{Theory vs Practice}
\end{figure}
By way of ascertaining the relative cost in time to perform the computations required the following measurements were obtained by timing the execution of commands to compress and decompress NZ\_vt03e\_069\_225750.lba.
\begin{table}[htdp]
\caption{Execution time}
\begin{center}
\begin{tabular}{|c|c|}\hline 
Algorithm & Time [s]\\ \hline
gzip & $17.3$ \\ \hline
gunzip & $6.4$ \\ \hline
bzip2 & $42.5$ \\ \hline
bunzip2 & $28.2$ \\ \hline
lzma & $318.8$ \\ \hline
unlzma & $24.7$ \\ \hline
\end{tabular}
\end{center}
\label{Compression results}
\end{table}%
%
%

\section{Conclusions and discussion}
The main conclusion of this work is that where the data flow from a radio telescope takes the form of a Gaussian distributed noise like signal and that ``memoryless'' conditions are assumed (probability of emission of any given symbol is independent of previously emitted symbols) then there is scope for compression. Compression is possible due to the non equiprobable emission of symbols inherent in the nature of the digitized Gaussian signal. The degree of compression available scales as an inverse function of the standard deviation of the Gaussian probability distribution.\\

It is of course another matter to conclude that the effort to compress is worthwhile. The ``costs'' of compression; time to compress and decompress, cost of hardware and power consumption must all be evaluated and weighed against the potential benefits.
\\

It would be desirable to be able to compress in real time, a compression algorithm that could be implemented in hardware as opposed to run as a sequential program on a compute engine would seem attractive from this point of view.
\\

If the strict memoryless condition assumed here does not hold then other forms of the Entropy function must be used;

\[H(A,P)=\sum_{i=1}^{N}P(a_i)\sum_{j=1}^{N}P(a_j | a_i)\log_2(P(a_j  | a_i)\] 
\\
($P(a_j | a_i)=$conditional probability for symbol $a_j$ to be emitted following symbol $a_i$) if there are (for instance) second order dependencies between symbols. In theory yet higher order dependencies are possible. Should any of these second or higher order dependencies exist then there is likely scope for higher levels of compression.

\end{document}